\documentclass[usenatbib]{mn2e}

\usepackage{graphics}
\usepackage{epsfig}
\usepackage{natbib}

\begin{document}

\title[Unusually HI-rich Galaxies]
{The Outer Stellar Populations and Environments of Unusually HI-rich Galaxies} 

\author [G.Kauffmann] {Guinevere Kauffmann \\
\thanks{E-mail: gamk@mpa-garching.mpg.de}
Max-Planck Institut f\"{u}r Astrophysik, 85741 Garching, Germany\\}

\maketitle

\begin{abstract} 
We investigate the  nature of HI-rich galaxies from the ALFALFA and GASS
surveys, which are defined as galaxies
in the top 10th percentile in atomic gas fraction at a given stellar mass.  
We analyze outer ($R>1.5 R_e$) stellar populations for a subset of face-on systems using 
optical $g-r$ versus $r-z$  colour/colour diagrams. 
The results are compared with
those from control samples that are defined without regard to 
atomic gas content, but are matched in redshift, stellar mass and
structural parameters. 
HI-rich early-type ($C>2.6$) and late-type ($C<2.6$) galaxies are studied separately. 
When compared to the control sample, the outer stellar populations 
of the majority of HI-rich early-type galaxies are shifted
in the colour/colour plane along a locus consistent with younger stellar ages, but similar metallicities.
The outer colours of HI-rich late-type galaxies are much bluer in $r-z$ than the HI-rich early types, 
and we infer that they have outer disks which are both younger
and more metal-poor.
We then proceed to analyze the galaxy environments of HI-rich galaxies on scales comparable to
the expected virial radii of their dark matter halos ($R < 500$ kpc). 
Low mass ($\log M_* < 10.5$)
HI-rich early-type galaxies have galaxy environments that differ significantly
from the control sample. HI-rich early types are more likely to be central
rather than satellite systems. Their satellites are also less massive and have
younger stellar populations. Similar, but weaker effects are found for
HI-rich late-type galaxies of the same mass. In addition, we find that the satellites of HI-rich late-types
exhibit a greater tendency to align along the major axis of the primary.
No environmental differences are found for massive ($\log M_* > 10.5$) HI-rich galaxies,
regardless of type.

\end{abstract}

\section{Introduction}

Recent targeted HI surveys of representative samples of galaxies selected by
stellar mass (Catinella et al 2010; 2013) have demonstrated that $L_*$ galaxies like
our own Milky Way span a broad range in atomic gas mass fraction.  The median HI
mass fraction for galaxies with stellar masses of 4-5 $\times 10^{10} M_{\odot}$
is $\sim 0.1$, but 10\% of galaxies in this mass range have gas fractions that
are more than three times higher (Kauffmann et al 2012; Lemonias et al 2013).

Galaxies that are significantly more gas-rich than average are extremely
interesting for further study. If the atomic gas is able to reach high enough
densities and metallicities to become shielded from the ambient UV radiation
field, it will form $H_2$ and from there, molecular clouds and stars.  Atomic
gas may thus be regarded as a ``reservoir'' for future star formation in a galaxy.
By studying HI-rich systems, we  gain significant insight into the way in
which galaxies grow in mass at the present day.

By comparing the colour profiles of large samples of HI-rich galaxies drawn from
the wide-field ALFALFA survey (Giovanelli et al 2005) to those of normal
spirals, we learned that gas-rich galaxies have unusually blue outer disks (Wang
et al 2011).  Follow-up spectroscopy  demonstrated that H$\alpha$ emission was
strong in these regions and that the ionized gas metallicities frequently
exhibited strong drops beyond the optical radius (Moran et al 2010; 2012). 
Analysis of the ratio of star formation rate to stellar mass surface densities as a function
of radius  demonstrated that most of the present-day growth in HI-rich
galaxies occurs in their outskirts. These results have recently been confirmed
by independent studies (Huang et al 2014).

We then undertook a follow-up program to map the HI in a small sample of 25
unusually HI-rich galaxies and 25 ``control'' galaxies with normal HI content
using the Westerbork Synthesis Radio Telescope (WSRT; Wang et al 2013, 2014).
The incidence of galaxies with disturbed HI morphologies was slightly lower in
the HI-rich sample than in the control sample, strongly arguing against a
scenario in which the extra HI gas had been recently added by a significant
merging event.  The main distinguishing characteristic of the HI-rich galaxies
was the very large radial extent of the HI in comparison to the optical light.
In addition, there was tentative evidence that the outer HI gas was more
clumpy/irregular in HI-rich galaxies than in the control sample.

In this paper, we extend our work on HI-rich galaxies by studying their outer
stellar populations in more detail, as well as their environments on scales of
20-500 kpc. We extract the largest available sample by combining the latest
public release from the ALFALFA survey (the alpha.40 catalog; Haynes et al 2011)
with the final data release from the GASS survey (Catinella et al 2013). As in
previous papers, we focus our attention on galaxies with stellar masses greater
than $10^{10} M_{\odot}$.  We divide this sample into two classes: a) Galaxies
that are structurally ``early-type'' in that they have have concentrated light
profiles, indicative of galaxies with large bulge-to-disk ratios, b) Galaxies
that are structurally ``late-type'' in that they have extended light profiles
indicative of disk-dominated systems. The stellar populations and environments
of the 10-25\%  most HI-rich systems are then compared with galaxies of the same
mass and structural type drawn from the full Sloan Digital Sky Survey data
release 7 (DR7; Abazajian et al 2009).

The paper is divided into two parts. In the first part, we address the question
of what can be learned about stellar populations using  available SDSS
photometry.  We discuss previous work on the colour profiles of disk galaxies
and ellipticals, and then show how unusually HI-rich galaxies of both types
compare with the parent samples. In the second part of the paper, we examine
whether HI-rich galaxies are more likely to be central or satellite galaxies in
their halos. We also analyze the stellar masses, star formation rates and
orientations of the galaxies in the vicinity of the HI-rich systems and analyze
whether any systematic difference is seen with respect to control samples.

\section {Stellar populations}

\subsection {Colour profiles of spiral and elliptical galaxies: background}

The existence of colour gradients in both spiral and elliptical  galaxies has
been known for more than 50 years (e.g. Tifft 1963). In the 1970's, astronomers
began to image galaxies at near-IR wavelengths for the first time. One of the important
motivations for near-IR surveys  was to break the degeneracy of the
UBV colour-system in distinguishing whether age or metallicity effects drive
basic scaling relations, for example the relation between galaxy colour and
velocity dispersion. Strom et al (1976) used the V-K radial colour
profiles of 5 elliptical and S0 galaxies to derive metallicity gradients and
compared them with the model predictions of Larson \& Tinsley (1974).  Aaronson, Huchra \& Mould( 1979)
and  Terndrup et al (1994)
emphasized the importance of dust extinction when interpreting radial colour trends in
galactic disks.

The most recent work on colour gradients in nearby galaxies has involved
statistical studies of much larger samples. The Sloan Digital Sky Survey (York
et al 2000) provided very uniform 5-band $u,g,r,i,z$ photometry for millions of
galaxies across a quarter of the sky and has served as the underlying data set
for most of these.  Tortora et al (2010) derived $g-i$ colour gradients for
50,000 galaxies with redshifts $z<0.05$. They found stronger gradients for low
mass galaxies than high mass galaxies. Suh et al (2010) studied $g-r$ colour
gradients for a large sample of early-type galaxies. They found that the
majority of early-types have flat colour profiles.  11 percent of the sample
exhibited negative colour gradients (i.e. bluer colours on the outside), while 4 percent
were ``blue core'' systems with positive gradients.  Gonzalez-Perez, Castander \& Kauffmann (2011)
showed that if galaxies were split into early and late-type classes at a
$r$-band concentration index on 2.6, there was no longer a significant
dependence of colour gradient on stellar mass for each type (i.e. the trend with
mass found by Tortora et al was driven by an increasing early-type fraction
among more massive galaxies).  The SDSS survey has also been used to study the
colour gradients of rare galaxy types. Roche, Bernardi \& Hyde (2010) compared the
colour profiles of brightest cluster galaxies with those of ordinary ellipticals
and SOs and found that they were flatter on average.  We note that most studies
of colour gradients using SDSS have been confined to a single colour (most often
$g-r$ or $g-i$).

\subsection {Two-colour profiles of HI-rich galaxies compared to the underlying population}

In figures 1-4, we show a series of $r-z$ versus $g-r$ colour-colour diagrams.
In each figure, the the 6 panels show the colours of galaxies at a fixed radius
normalized to the radius $R_e$, defined as the radius containing half the total
r-band luminosity of the galaxy. The radii plotted range from one tenth $R_e$ to
2.5 $R_e$. The SDSS photometry is not deep enough to extract $r-z$
colours at radii larger than this for individual galaxies.

Figure 1 shows results for early-type galaxies in the mass range $10.0<\log
M_*<10.3$, while Figure 2 shows results for late-type galaxies in the same mass
range.  We divide galaxies into early/late-type using a simple cut on the
$r$-band concentration index $C$ of 2.6, where $C$ is defined as the ratio of
the radii enclosing 90\% and 50\% of the total $r-$band light from the galaxy.
Shimasaku et al (2001) have shown that this cut can separate galaxies that are visually
classified as E/S0/Sa from later-type spirals and irregulars with reasonable
accuracy. Figures 3 and 4 are the same as Figures 1 and 2, except that they show  galaxies
in the mass range $10.3<\log M_*<10.6$.

In each figure, the grey-scale contours show the location of
the general population of galaxies  in $r-z$ versus $g-r$
colour/colour space. These contour plots
are constructed as follows: \begin {enumerate} \item
From the data release 7 (DR7) of the Sloan Digital Sky Survey,
we select all
spectroscopically-targeted galaxies with redshifts in the range $0.015<z<0.06$
and with stellar masses greater than $10^{10} M_{\odot}$. The stellar masses are taken
from the the MPA/JHU value-added
catalogue (http://www.mpa-garching.mpg.de/SDSS/DR7/). We discard galaxies that
are highly inclined with axial ratios $b/a < 0.6$ in order to minimize the
effects of dust extinction on our analysis.  \item The SDSS photometric pipeline
extracts azimuthally averaged radial surface brightness profiles for all the
objects in the survey.  In the catalogs, this is given as the average surface
brightness in a series of annuli with fixed angular dimension, which we then
interpolate onto a fixed grid in units of $R/R_e$. We discard the outer annuli
in which the errors on the $g-r$ or $r-z$ colours are greater than 0.2. \footnote {This corresponds to a
20\% error on the colour, which may seem rather large. We note that will mostly be concerned with 
{\em relative comparisons} between different sub-samples at fixed radius in this section. In section 2.4,
where we attempt to interpret the colours in terms of physical
parameters, we will create higher $S/N$ estimates of outer colours.}   \item
The contours indicate the stellar mass-weighted fraction of the galaxy
population with $r-z$ and $g-r$ colours in a given range. We bin the $r-z$
versus $g-r$ colour/colour plane into cells of size $0.05\times0.05$.  Each galaxy is
weighted by $M_*$, its mass, and $1/V_{max}$, the inverse of the volume  over which
it is found in our sample.  Our chosen contours are spaced a factor 2.5 apart in
mass fraction, i.e. the lowest  (black) contour corresponds to a factor 244
lower mass-weighted fraction than the highest (white) contour.  \end {enumerate}

\begin{figure*}
\includegraphics[width=125mm]{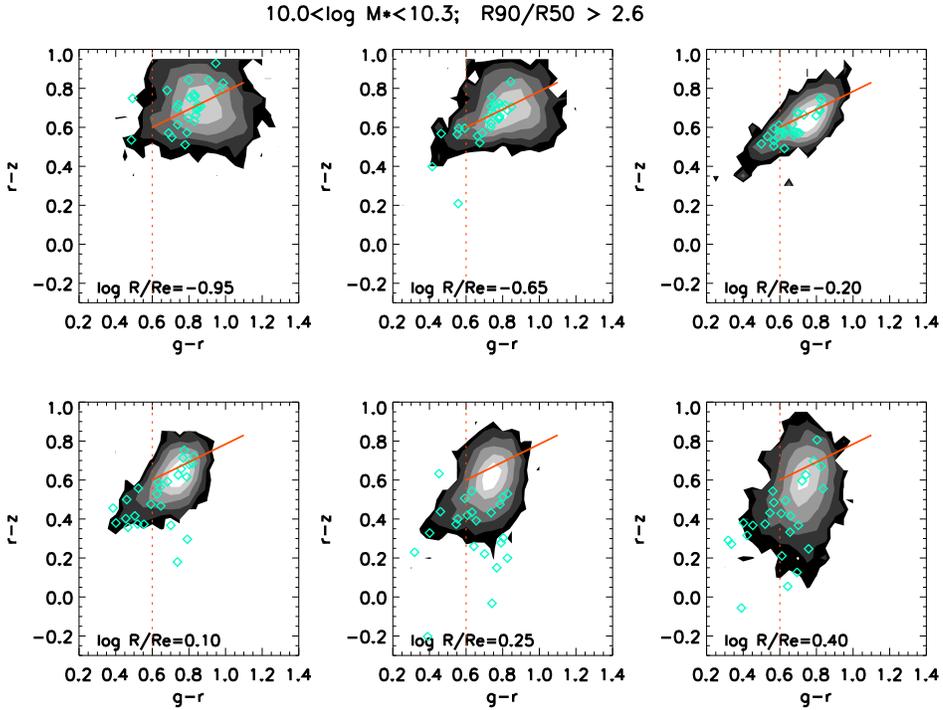}
\caption{ The distribution of the stellar populations of galaxies  in the $r-z$ versus $g-r$ colour/colour  
plane in 6 radial bins. Results are shown early-type for galaxies in the stellar mass range
$10 < \log M_* < 10.3$ and with concentration index $R_{90}/R_{50}>2.6$  
The contour plots show the stellar mass and $1/V_{max}$ weighted fractional distribution of
all galaxies in the SDSS DR7. The contour levels are spaced a factor of 2.5  apart and range
from 0.05 (white) to 0.0002 (black). The cyan diamonds show galaxies from the ALFALFA
and COLD GASS surveys which are in the upper 10th percentile in HI mass fraction. 
The dashed and solid red lines are placed in the same location in each panel to guide the eye.   
\label{models}}
\end{figure*}

\begin{figure*}
\includegraphics[width=125mm]{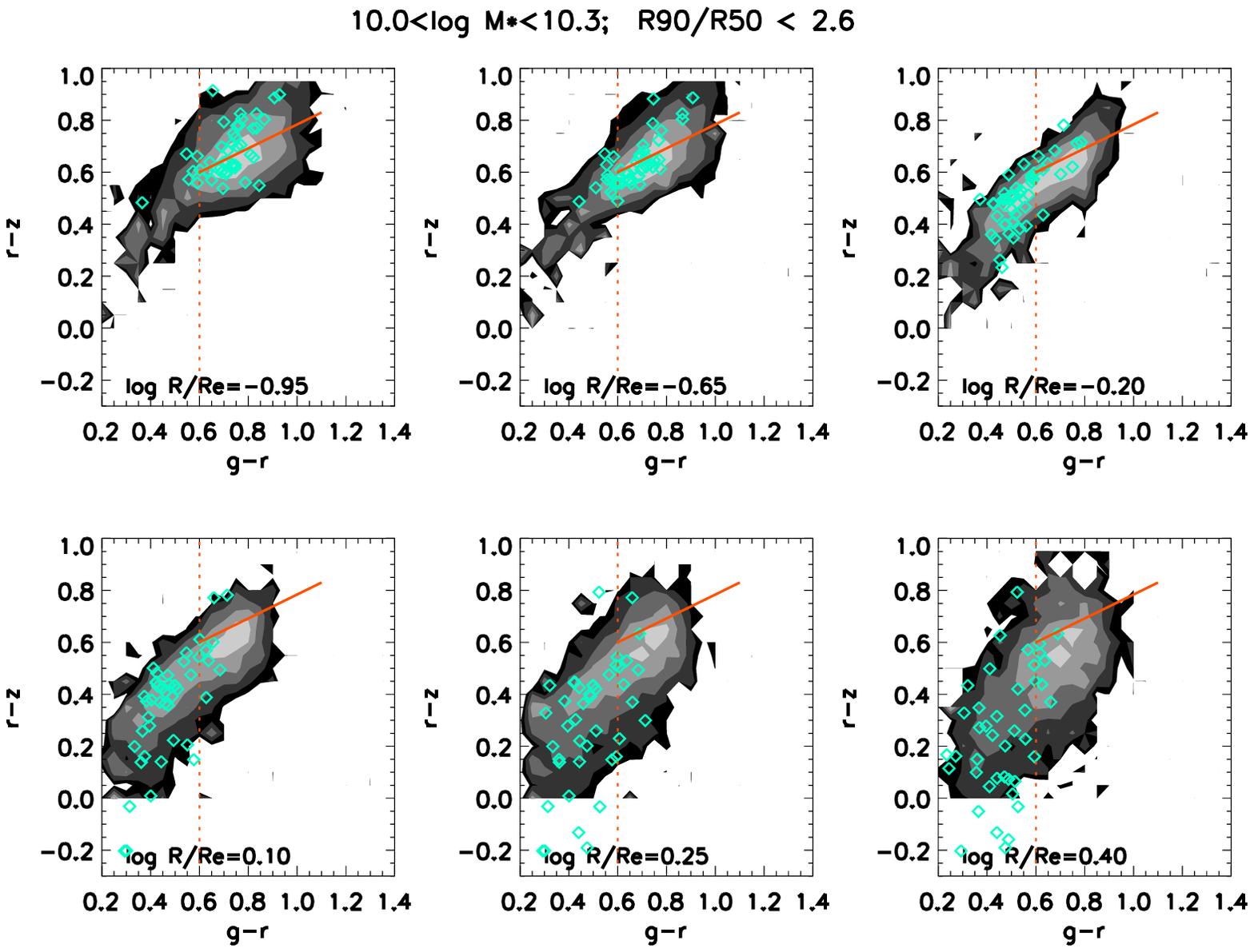}
\caption{ As in Figure 1, except for galaxies in the stellar mass range    
$10 < \log M_* < 10.3$ and with concentration index $R_{90}/R_{50}<2.6$  
\label{models}}
\end{figure*}

\begin{figure*}
\includegraphics[width=125mm]{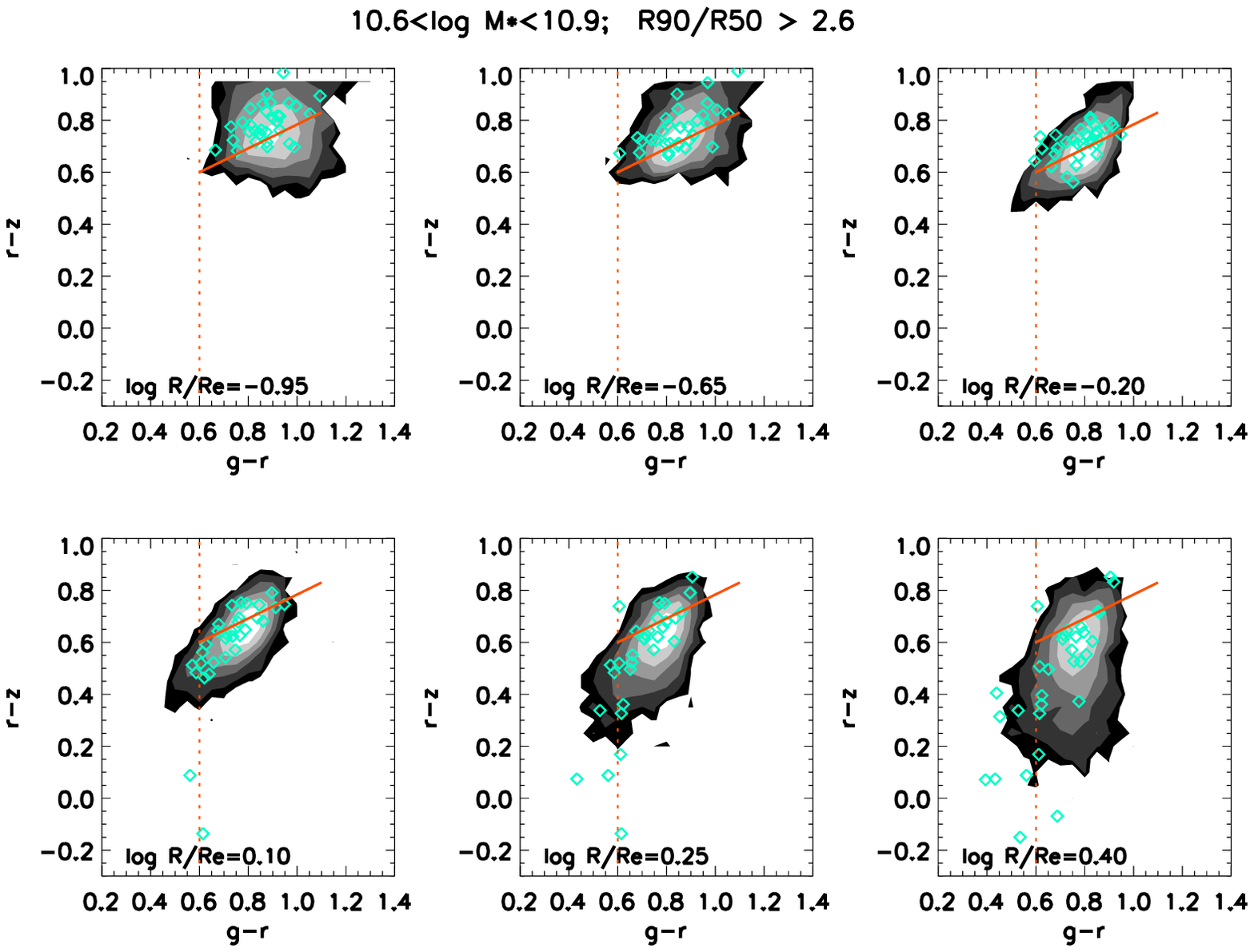}
\caption{ As in Figure 1, except for galaxies in the stellar mass range    
$10.6 < \log M_* < 10.9$ and with concentration index $R_{90}/R_{50}>2.6$  
\label{models}}
\end{figure*}

\begin{figure*}
\includegraphics[width=125mm]{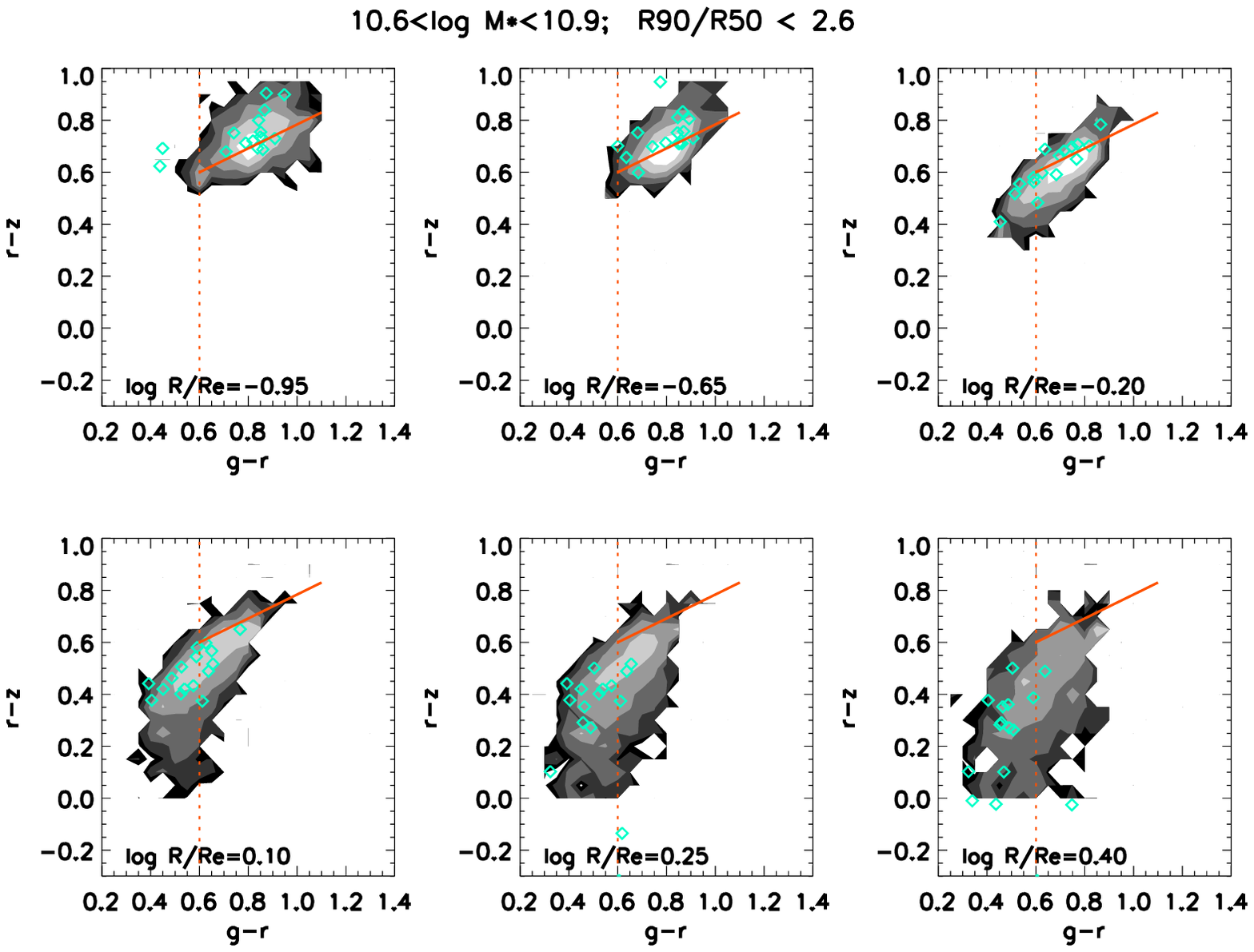}
\caption{ As in Figure 1, except for galaxies in the stellar mass range    
$10.6 < \log M_* < 10.9$ and with concentration index $R_{90}/R_{50}<2.6$  
\label{models}}
\end{figure*}

The main  two-colour trends can be summarized as follows:  

On average, early-type galaxies exhibit weak gradients in both $g-r$ and $r-z$
colours. The peak of the distribution shifts from $g-r= 0.87$ at 0.1$R_e$ to
$g-r=0.7$ at 2.5$R_e$, while $r-z$ changes from 0.7 to 0.6.  As noted by Suh et
al (2010), a minority of early-type galaxies have significantly bluer outer
regions. In our plots, it can be seen that the blue tail is much more
prominent in $r-z$ than it is in $g-r$, but as we will show later, at least part  of this 
is due to larger errors on the $r-z$ colours at large radii.

Comparison of Figures 1 and 3 shows that more massive early type galaxies
exhibit less scatter in their central colours and have weaker colour gradients.
The shift in $g-r$ colour from 0.1$R_e$ to 2.5$R_e$ is  0.1 for $10^{10}
M_{\odot}$ ellipticals compared to 0.17 for ellipticals that are 4 times more
massive. However,  $r-z$ colour gradients do not appear to vary with stellar
mass.  

As seen in Figure 2, the central colours of late-type galaxies are spread much
more widely than those of early-type galaxies.  As radius increases, the mass
weighted fraction of galaxies in the very blue part of the $r-z$ versus $g-r$
colour plane increases. Interestingly, however, the location of the {\em peak}
of the colour-colour distribution remains very nearly fixed with radius.

Comparison of Figures 2 and 4 show that  massive late-type galaxies exhibit
significantly smaller dispersion in colour in their central regions than their
less massive counterparts.  However, $g-r$ and $r-z$ colour gradients  appear to
be {\em stronger} for more massive late type galaxies. At radii $R> 1.5 R_e$,
the $g-r$ colours of low mass late-types exhibit a  peak in the range 0.7-0.8,
i.e. at relatively red colours.  This red peak is not found in the high mass
late-type sample. Likewise, there are no massive late-type galaxies with $r-z$
colours greater than 0.65 in their very outer regions, whereas a fair number
exist in the low mass sample.

Cyan symbols in Figures 1-4 show the locations of HI-rich galaxies in $r-z$
versus $g-r$ colour-colour space. Our HI-rich sample is constructed by combining
the final data release of the GALEX Arecibo SDSS Survey (GASS; Catinella et al
2013) with the  alpha.40 HI Source Catalog from the Arecibo Legacy Fast ALFA
Survey (Haynes et al 2011). GASS is designed to reach a limiting HI mass
fraction of between 1.5 and 3\% for galaxies with $M_* > 10^{10} M_{\odot}$ and
thus can be used to define our HI-rich sample.  In Figure 5, we plot plot 50th,
75th, and 90th percentiles of the distribution of M(HI)/$M_*$ as a function of
stellar mass for early-type galaxies with concentration index $C>2.6$ and for
late-type galaxies with $C<2.6$.  This plot is constructed using the GASS DR3
``representative sample'', which samples the M(HI)/M$_*$ distribution in  each
stellar mass in an unbiased way (see Catinella 2010; 2013 for details). Note that
more than half of early-type galaxies with masses greater than $10^{11}
M_{\odot}$ are not detected in HI. For
these galaxies, the median value of M(HI)/$M_*$ is plotted at the value of the
upper limit, correpsonding to a HI mass fraction of 0.015. 
This is an  over-estimate of the true median value.  This will
not concern us here, because we define HI-rich galaxies as those with HI mass
fractions that lie above the 90th percentile points shown as filled cyan circles
in the two panels. As can be seen, the 90th percentile points span gas fractions
in the range 0.1-0.3 for early-type galaxies and 0.3-1 for late-type galaxies.

\begin{figure}
\includegraphics[width=88mm]{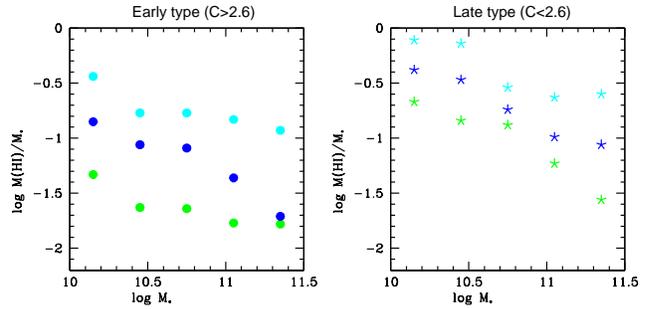}
\caption{ The 50th (green symbols), 75th (blue symbols) and 90th (cyan symbols) percentiles of                
the distribution of M(HI)/M$_*$ are plotted as a function of the logarithm
of the stellar mass for early-type galaxies with $C>2.6$ (left) and for late-type
galaxies with  $C<2.6$ (right).
\label{models}}
\end{figure}

It is immediately clear that HI-rich galaxies are almost always displaced
towards the blue region of colour-space when compared to the underlying
population. A closer examination of Figures 1-4 reveals, however, that the
nature of the displacement depends  on radius, on galaxy mass, and on morphological
type.  HI-rich early-type galaxies with low masses 
have  bluer $g-r$ colours in their inner regions ($R< R_{50}$), but this is not
seen for  HI-rich early-type galaxies with high masses.      
Both low and high mass HI-rich late-type galaxies become significantly bluer than the underlying population in $g-r$ at
$R>0.5R_{50}$. The $r-z$ colours of both early-type and late-type HI-rich galaxies with low stellar
masses  are not noticeably displaced from
the underlying population at $R< R_{50}$. At larger radii, there is increasing scatter in  $r-z$ colour towards the blue.
In section 2.4, we will come back to whether this effect is real or simply a reflection of
increasing photometric errors in the lower surface brightness outer
regions of the galaxy.  Finally, we note that   
the $r-z$ colours of both early-type and late-type HI-rich galaxies with high stellar masses
are  displaced to slightly {\em redder} values in their central regions. Once again, we find  large
scatter to bluer $r-z$ colours at large radii. 

\subsection {Interpretation of $r-z$ versus $g-r$  colour/colour diagrams in terms of
stellar population parameters}

The combination of optical and optical/infrared colours has been used a way of
estimating stellar ages and metallicities in both spiral and elliptical galaxies
(Bell \& De Jong 2000; MacArthur et al 2004; Tortora et al 2010). These methods
are only likely to yield accurate answers if the star formation histories of
galaxies have been smooth rather than bursty, and if there is very little dust.
These limitations are illustrated in detail in Figure 6, where we show the
predicted locations of galaxies with different star formation histories in the
$r-z$ versus $g-r$ colour-colour plane using the population synthesis models of
Bruzual \& Charlot (2003).

The coloured tracks indicate the location of galaxies that have had smooth star
formation histories of the form SFR $\propto e^{-t/\tau}$, where $\tau$
ranges from 0.1 to 100 Gyr. Three different look-back times for star formation
to commence are adopted: 13, 6  and 3 Gyr. Different colours indicate different
metallicities: magenta corresponds to 1.5 solar, green to 0.5 solar and blue to
0.1 solar.  As can be seen, models of differing age, but the same metallicity,
occupy very narrow loci in the $r-z$ versus $g-r$ plane. Under the assumption
of smooth star formation histories and no dust, galaxies with different stellar
metallicities can be separated quite easily if their $r-z$ colours can be
measured accurately enough.

If galaxies have experienced bursts of star formation in their recent past, the
situation becomes significantly more complicated. The four different panels in
Figure 6 show four different possibilities. In the top-left panel we show what
happens if a metal-rich burst is superposed on a smooth metal-poor $\tau$ model,
while the bottom left panel shows the opposite case of a metal-poor burst
superposed on a metal-rich $\tau$ model.  The top and bottom-right panels show
the cases of metal-poor burst with metal-poor $\tau$, and metal-rich burst with
metal-rich $\tau$, respectively. In all cases, ``metal-rich'' corresponds to
metallicities between solar and 1.5 solar, while ``metal-poor'' corresponds to
metallicities between 0.1 and 0.25 solar. The bursts are allowed to begin at
look-back times ranging from 2 Gyr to 0.1 Gyr past and last for 0.1 Gyr.  Between
1\% and 90\% of the total mass of the galaxy is allowed to form in the burst.

As can be seen, the power of the $r-z$ colour to differentiate metallicity can
be compromised by bursts. Superposing a metal-rich starburst on a metal-rich
$\tau$ model can lead to $g-r$ and $r-z$ colours that overlap the locus of
$\tau$ models with 0.1 solar metallicity.  Our experiment of superposing
different model star formation histories leads to the following general
conclusion: if, in the absence of dust, the $r-z$ colour of a galaxy is red,
then the galaxy is  metal-rich.  However, the opposite does not necessarily
hold. If the $r-z$ colour is blue, then the galaxy may be metal poor, or it may
have undergone a recent burst of star formation.

The effect of dust on the $r-z$ versus $g-r$ colours is illustrated by the red
vector plotted in the bottom-right corner of the top-right panel in Figure 6. We
adopt a standard Milky Way extinction law (Cardelli, Clayton \& Mathis  1989) and
the length of the vector corresponds to $\tau$(V)=0.5 mag and R(V)=3.1.
\footnote {The choice of extinction law makes almost no difference to the
orientation of the vector in the $r-z$ versus $g-r$ plane}. Extinction will 
move galaxies away from the loci defined by the dust-free $\tau$ models in the
sense of causing them to have redder-than-predicted $r-z$ colours at a given
metallicity.

\begin{figure}
\includegraphics[width=88mm]{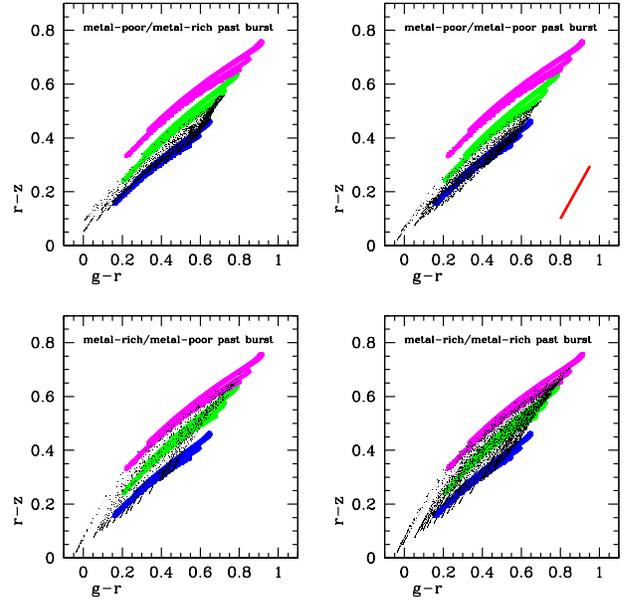}
\caption{
The predicted locations of galaxies with different star formation histories in the
$r-z$ versus $g-r$ colour-colour plane. The coloured tracks indicate the location of galaxies that have had smooth star
formation histories. Different colours indicate different
metallicities: magenta corresponds to 1.5 solar, green to 0.5 solar and blue to
0.1 solar. The black points indicate galaxies that have experienced past bursts (see text).
The effect of dust is illustrated by the red vector.                  
\label{models}}
\end{figure}

\subsection {Application to outer stellar populations}  

Given the complications discussed in the previous section, we do not attempt a
detailed quantitative analysis of age or metallicity trends as
a function of radius in galaxies in this paper.
However, the strong trend  towards  bluer outer $g-r$ and $r-z$ colours in the galaxy
population as a whole, and in in HI-rich galaxies in particular, is worthy of
further analysis. The outer regions of galaxies viewed face-on are sites where
one might reasonably conjecture that neither recent bursts of star formation nor
dust play a significant role.

One question that arises is the extent to which the ``blue tails" seen in the 
outer radii in Figures 1-4  are simply due to increasing errors in
the measured colours at large radii. In particular the SDSS $z$-band photometry has lower
signal-to-noise than the $g-$ and $r$-band photometry, so the $r-z$ colours 
will exhibit greater spread simply due to photometric errors. To answer this, we construct a higher
$S/N$ estimate of outer colour for each galaxy by binning together the light
from radii larger than $1.5 R_{50}$. We select the subset of galaxies with
errors on both their  $g-r$ and $r-z$ outer colours of less than 0.04 and plot
these in Figure 7 on top of our grid of $\tau$ models. The left panels show
results for early-type galaxies, while the right panels show results for
late-type galaxies. We have added a yellow track, indicating the location of
single stellar population (SSP) models of 0.01 solar metallicity. The yellow
track may be regarded  as a  theoretical lower limit to the allowed range of
$r-z$ colours at a given value of $g-r$.  \footnote{We have compared the low
metallicity SSPs from the Bruzual \& Charlot (2003), Maraston (2005) and Conroy, Gunn \& White (2009)
models and find that they all yield comparable results.} 
As can be seen, even
though the 0.01 solar SSP models are very blue in $g-r$,  there is almost no
difference in $r-z$ at fixed $g-r$ colour compared to the 0.1 solar $\tau$
models. This means that stellar populations with  metallicities below 0.1 solar
can no longer be clearly distinguished in this diagram.

The red dashed lines in the top panels of Figure 7 show our chosen demarcation
line for galaxies that are ``unusually blue'' in outer $r-z$ colour. Our
demarcation line is chosen to pass through the 0.1 solar $\tau$ model track.
Galaxies lying below this line are plotted as cyan coloured points, while those
lying above the line are plotted as black points. Around 10\% of both early-type
and late-type galaxies are included in the ``unusually blue'' sub-sample.  In
the bottom panels, we plot the same galaxies in the plane of 4000 \AA\ break
strength (D$_n$(4000)) versus stellar surface mass density.  The 4000 \AA\ break
strength is a relatively clean  measure of the age of the central stellar
population of the galaxy within the 3 arcsecond diameter fibre aperture. Because
the index is defined over a very narrow range in wavelength, it is insensitive
to dust extinction. The stellar surface mass density is defined as $0.5 M_*/\pi
R_{50}^2$, where $R_{50}$ is the half-light radius of the galaxy. As can be
seen, early-type galaxies with unusually blue outer $r-z$ colours have central
stellar population ages and densities that are indistinguishable from the
underlying population. Late-type galaxies with unusually blue outer colours have
similar central ages to the underlying population, but are shifted to lower mass
densities by a factor of 3 (implying that at fixed stellar mass, their
half-light radii are a factor of 1.7 larger).  These differences are quantified
in more detail in Figure 8, in which we show histograms of D$_n$(4000) and log
$\mu_*$ for the two samples.

\begin{figure}
\includegraphics[width=88mm]{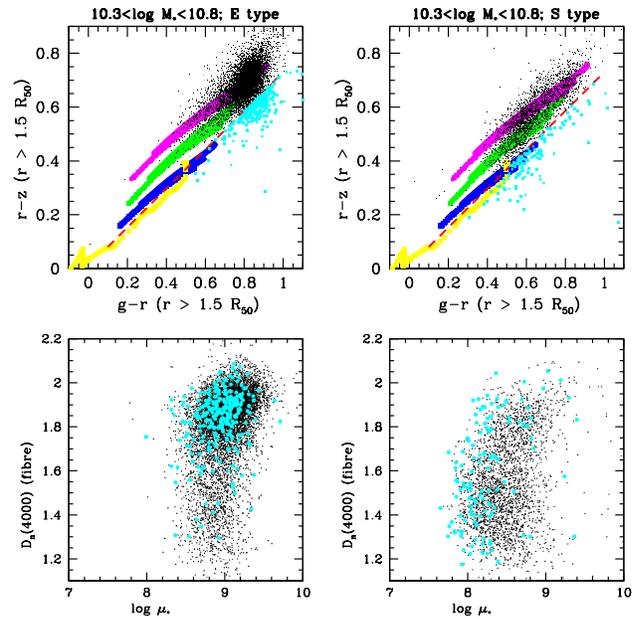}
\caption
{{\bf Top:} The outer ($R>1.5 R_e$) colours of galaxies are plotted 
in the $r-z$ versus $g-r$ colour-colour plane. We have selected galaxies with
errors in both $g-r$ and $r-z$ outer colours of less than 0.04 in the stellar mass
range $10.3 < \log M_* < 10.8$. Results are shown separately for
early-type ($C>2.6$) and late-type ($C<2.6$) galaxies.  
The coloured tracks indicate the location of galaxies that have had smooth star
formation histories. Different colours indicate different
metallicities: magenta corresponds to 1.5 solar, green to 0.5 solar and blue to
0.1 solar. The yellow track shows the locus of 0.01 solar metallicity
simple stellar populations (SSPs).
Galaxies above the red-dashed line are plotted in black; galaxies below the
red-dashed line are plotted in cyan.
{\bf Bottom:} We show the same galaxies in the top panels in the
plane of 4000 \AA\ break strength (D$_n$(4000)) versus the logarithm of
the stellar surface mass density. Colour-coding of points is as in the
top panels.
\label{models}}
\end{figure}
 
\begin{figure}
\includegraphics[width=88mm]{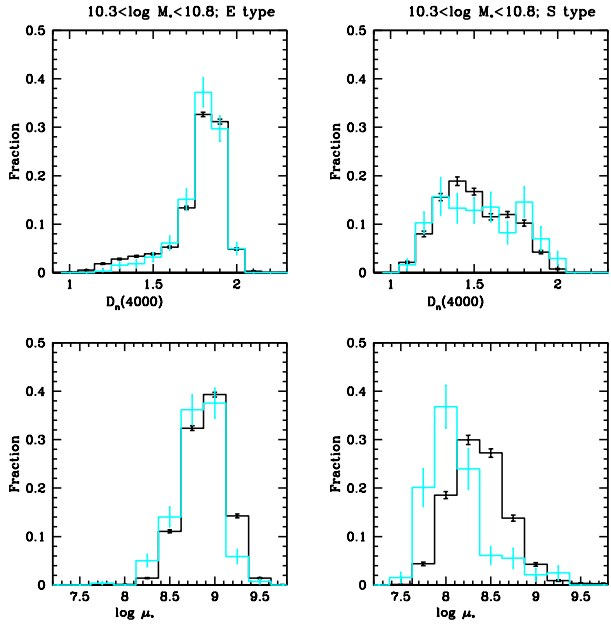}
\caption
{Histograms of 4000\AA\ break strength and stellar surface mass density for the
same galaxies plotted in Figure 7. Black histograms are for the galaxies
lying above the dashed red line in Figure 7, while cyan histograms are for
the galaxies lying below this line. Errorbars are computed by
bootstrap resampling the samples.
\label{models}}
\end{figure}
 
If we interpret galaxies lying below the red dashed lines in Figure 7 as having
metal-poor outer-stellar populations, then the results in Figure 8 show that
early-type galaxies with metal-poor outer stellar populations have central
stellar mass densities and ages that are indistinguishable from the
early-type galaxy population as a whole. Late-type 
galaxies with metal-poor outer stellar population also have central
stellar population ages that are  indistinguishable from other
late-type galaxies, but they have significantly lower stellar densities
than the parent population.       

Catinella et al (2010) show that the HI mass fraction in galaxies is strongly correlated 
with stellar density in addition to UV/optical colour. In Figure 9, we plot the 
outer ($R > 1.5 R_{50}$) $r-z$ versus $g-r$  colours of HI-rich galaxies on top of the
parent population. To increase our HI-rich sample statistics, we include
all galaxies in the upper 25th percentile of HI gas mass fraction and with colour errors less than 0.08.
 The blue points indicate the locus of
continuous models with 0.1 solar metallicity for reference. Results are shown for two
stellar mass ranges: $10.0< \log M_* < 10.5$ and  $10.5< \log M_* < 11.0$. 

As can be seen, the
outer stellar populations of HI-rich early-type galaxies in the lower mass bin lie along a track in the
$r-z$ versus $g-r$ colour/colour diagram  consistent with younger
ages but  metallicities that are the same as the
outer stellar populations of the  parent population. We infer that {\em if} the extra HI gas in these
systems is of external origin, it must have been accreted at high metallicity. 
We also note that there are only a few HI-rich early-type systems that overlap
the unusually  blue  $r-z$ population  plotted in cyan in the left panels of Figures 7 and 8. 
The early-type galaxies with the bluest outer $r-z$ colours in Figures 1 and 3 have 
have errors on their $R>1.5R_{50}$ colour measurements that are too large to
allow them to be included in this sample.
In contrast, 
the outer stellar populations of low-mass HI-rich late-type galaxies are shifted to
both younger ages and lower metallicities. The shift in metallicity is largest 
for the HI-rich galaxies with the youngest outer disks. The most extreme amongst
these systems have estimated stellar metallicities at $R > 1.5 R_{50}$ of a tenth
solar or less. \footnote {We note that the scatter below the tenth solar track is
consistent with the typical error on the $r-z$ outer colour}       
The shifts in colour for the high mass HI-rich galaxies are very much weaker. In particular,
the high-mass HI-rich early-types display no colour offsets at all.

\begin{figure*}
\includegraphics[width=128mm]{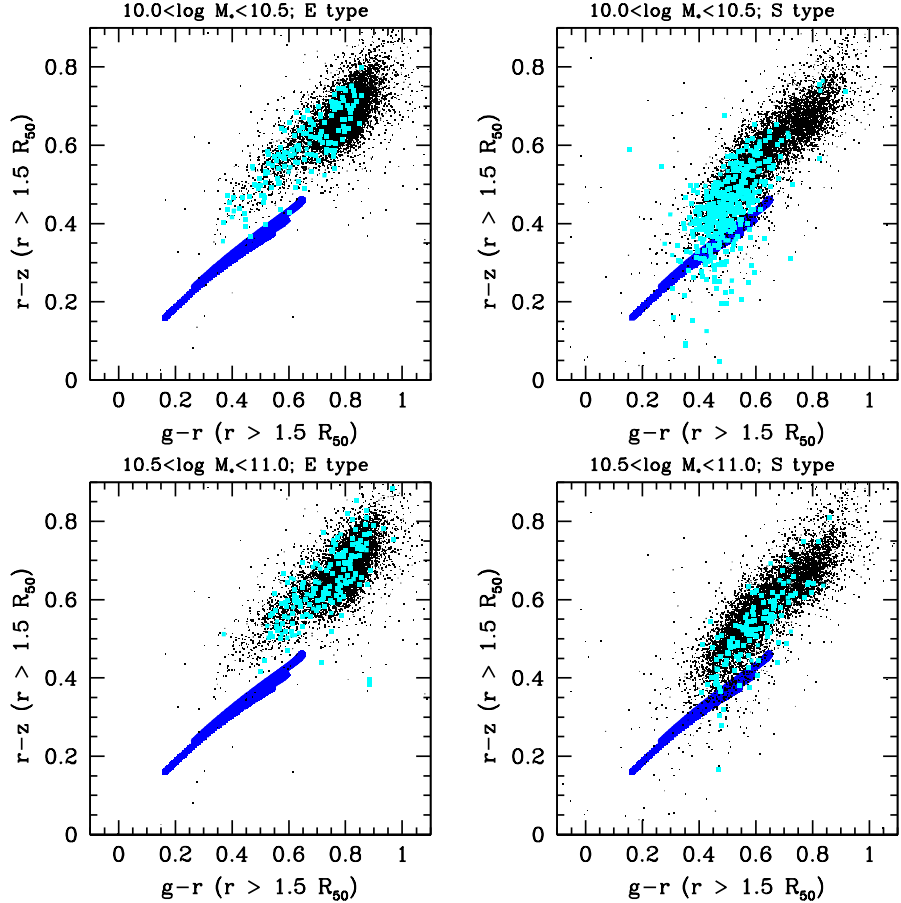}
\caption
{The outer ($R>1.5 R_e$) colours of galaxies are plotted
in the $r-z$ versus $g-r$ colour-colour plane. We have selected galaxies with
errors in both $g-r$ and $r-z$ outer colours of less than 0.04 in the stellar mass
ranges $10.0 < log M_* < 10.5$ and  $10.5 < log M_* < 11.0$. Results are shown separately for
early-type ($C>2.6$) and late-type ($C<2.6$) galaxies. 
The blue track indicates the location of galaxies of 0.1 solar
metallicity that have had smooth star
formation histories. Cyan points indicate galaxies with HI gas mass fractions
in the upper 25th percentile for their stellar mass detected in the COLD GASS
or ALFALFA surveys. We only plot galaxies with
colour errors less than 0.08.
\label{models}}
\end{figure*}

\section {Dark matter halo-scale environments of unusually HI-rich galaxies}

In this section we compare the environments of HI-rich galaxies to the
underlying population.  We focus on scales less than 500 kpc, roughly
corresponding to the predicted virial radii of the dark matter halos that host
$L_*$ galaxies.

The influence of environment on atomic gas in galaxies has been studied in the
past using two main methods: 1) In wide-field surveys such as HIPASS (Barnes et
al 2001) or ALFALFA (Giovanelli et al 2005), it is possible to compute the
auto-correlation function of HI-selected galaxies and compare this to the
auto-correlation function of optically-selected galaxies (Passmoor et al 2011;
Martin et al 2012; Papastergis et al 2013).  These studies find that HI-selected
galaxies cluster very weakly, with a clustering length $r_0$= 3.3 h$^{-1}$ Mpc,
and are more strongly anti-biased with respect to the dark matter distribution
on small scales compared to large scales. 2) There have also been studies
focusing on the HI properties of galaxies in  clusters (Giovanelli \& Haynes
1985; Kenney \& Young 1989; Solanes et al 2001; Chung et al 2009), showing that
HI has been removed from the disks of spirals in the central regions of clusters
by processes such as ram-pressure stripping or tidal interactions.  In recent
work, Catinella et al (2013) cross-correlated  GASS galaxies with the updated SDSS
galaxy group catalogue of Yang et al (2007) and  showed that galaxies in halos with
estimated virial masses greater than $10^{13} M_{\odot}$ are gas-deficient
compared to galaxies in lower mass halos.

We note that these past studies have almost exclusively probed physical
processes leading to the {\em removal} of atomic gas from galaxies in dense
environments such as clusters and groups.  This is true also for the correlation
function studies, because the amplitude on small scales  scales ($< 1$ Mpc) is
simply proportional to  the square of the galaxy density within individual dark
matter halos.  In this paper, we take a different approach and study environment
in a way that is designed to probe possible {\em gas accretion processes} in
HI-rich galaxies. Our analysis focuses on whether HI-rich galaxies of different
stellar masses and morphological types are more likely to be central or
satellite systems within their dark matter halo. We also  analyze not only the
number of satellites around HI-rich galaxies, but also properties such as    
their stellar masses and the ages of their stellar populations. Finally, we look
at whether the satellites of HI-rich galaxies display any preferred
orientation with respect to the major axis of the primary (the so-called
Holmberg effect).

\subsection {Central and satellite properties}

In a redshift survey, there is no completely accurate way of separating galaxies
into those that reside at the centers of their dark matter halos (so-called
central galaxies) and those that are satellite systems gravitationally bound to
a more massive central object. We adopt a simple definition used in previous analyses
(e.g Kauffmann et al 2012) whereby a galaxy is classified as central if it has
stellar mass larger than all its neighbours within a projected radius of 500 kpc
with velocity difference $\Delta V <$ 500 km/s. It is also classified as  a
central if no neighbours are found within this volume. A galaxy is classified as
a satellite if it has a more massive neighbour within $\Delta R < 500$ kpc and
$\Delta V <$ 500 km/s.  In this paper, we add an addition class of  ``binary''
galaxies, defined that those with a neighbour within a factor of two of
their own stellar mass.

Our HI-rich sample is constructed as described in the previous section, except
that for the environmental analysis we include all galaxies regardless of their
inclination.  For every HI-rich galaxy, we select a sample of 10 ``control''
galaxies, randomly drawn from the DR7 spectroscopic catalogue. The control
galaxies are chosen to match the HI-rich galaxy to within 0.2 dex in log $M_*$,
0.2 dex in log $\mu_*$, 0.01 in redshift, and 0.2 in axial ratio.  The
environmental properties of the HI-rich sample are then analyzed relative to the
control sample, which is selected without regard to atomic gas fraction.

In Figures 10 and 11, we show results for early and late-type HI-rich and control
galaxies with stellar masses in the range $10.0<\log M_*<10.3$. In the top
panels, we plot the fraction of galaxies classified as central/satellite/binary
and the distribution of the number of neighbours with $\Delta R < 500$ kpc and
$\Delta V < 500$ km/s. In the bottom panels we plot the stellar mass and 4000
\AA\ break strength distributions of the satellites. Black histograms show
results for the HI-rich galaxies, while red points show results for the control
sample, which is 10 times larger in size. Error bars are computed by boot-strap
resampling both samples.

Interestingly, environmental differences between HI-rich galaxies and the
control sample are much more apparent for early-type galaxies than for late-type
galaxies.  The fraction of HI-rich early-type galaxies that are centrals is
0.75, compared to 0.55 for the control sample. Their satellite populations
consists of lower mass galaxies with younger stellar populations. Similar
differences are found  for HI-rich spirals, but the effect is somewhat weaker.

Figure 12 and 13 are the same as figures 10 and 11, except for HI-rich and
control galaxies with stellar masses in the range $10.6<\log M_*<10.9$. Here we
see no difference between the environments of HI-rich early-type galaxies and
the control sample. There is a weak tendency for the satellites of massive
HI-rich spirals to have younger stellar populations compared to the control
sample, but otherwise the two samples are identical in terms of
central/satellite fractions and the number and stellar mass distributions of
neighbours.

\begin{figure}
\includegraphics[width=88mm]{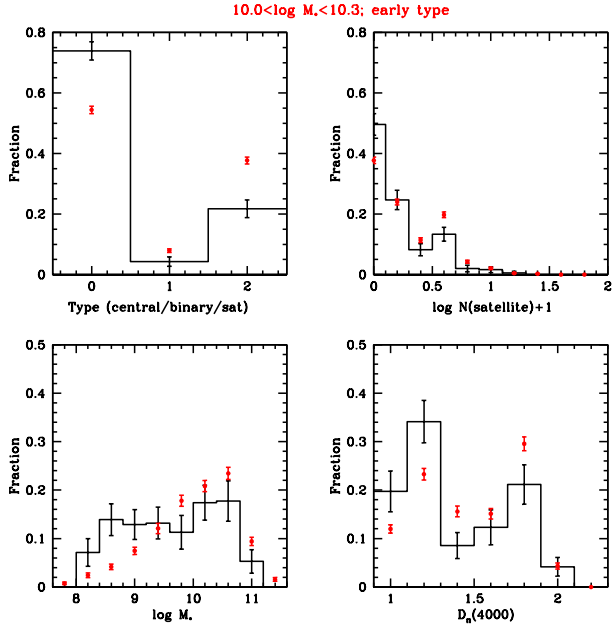}
\caption
{Distribution of central/satellite/binary galaxy fractions (upper left),
satellite numbers (upper right), satellite stellar masses (lower left)
and satellite 4000 \AA\ break strengths. Results are shown for
early-type ($C>2.6$) galaxies in the stellar mass range $10<\log M_*<10.3$.
Black histograms show results for HI-rich galaxies with HI mass fractions
in the upper 10th percentile, while red symbols show
results for control samples (see text). Errorbars have been computed by
bootstrap resampling. Note that in the top left panel, Type=1,2,3
refers to central, satellite, and binary, respectively.
\label{models}}
\end{figure}

\begin{figure}
\includegraphics[width=88mm]{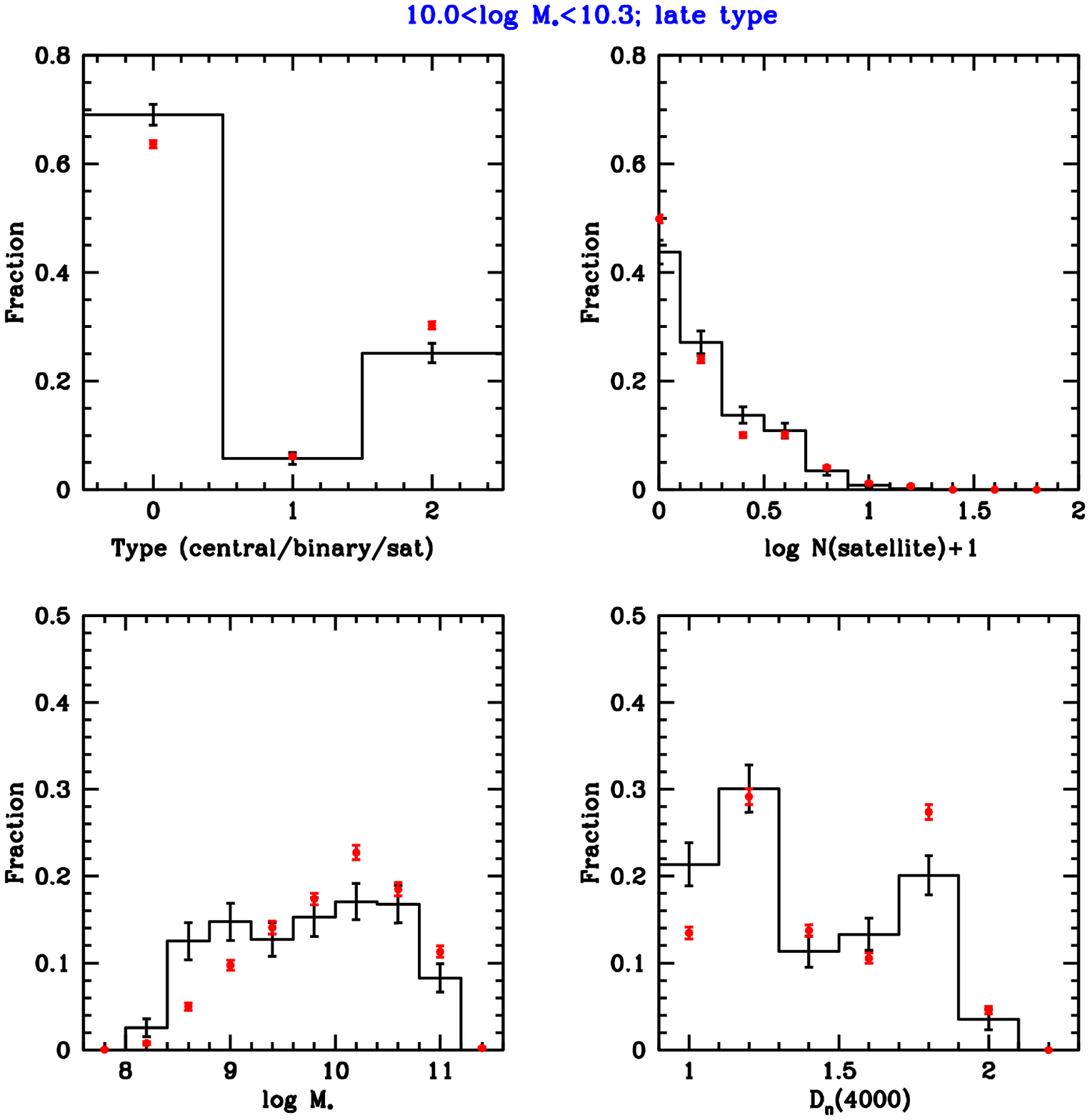}
\caption
{As in Figure 10, except for                                            
late-type ($C<2.6$) galaxies in the stellar mass range $10<\log M_*<10.3$.
\label{models}}
\end{figure}
 
\begin{figure}
\includegraphics[width=88mm]{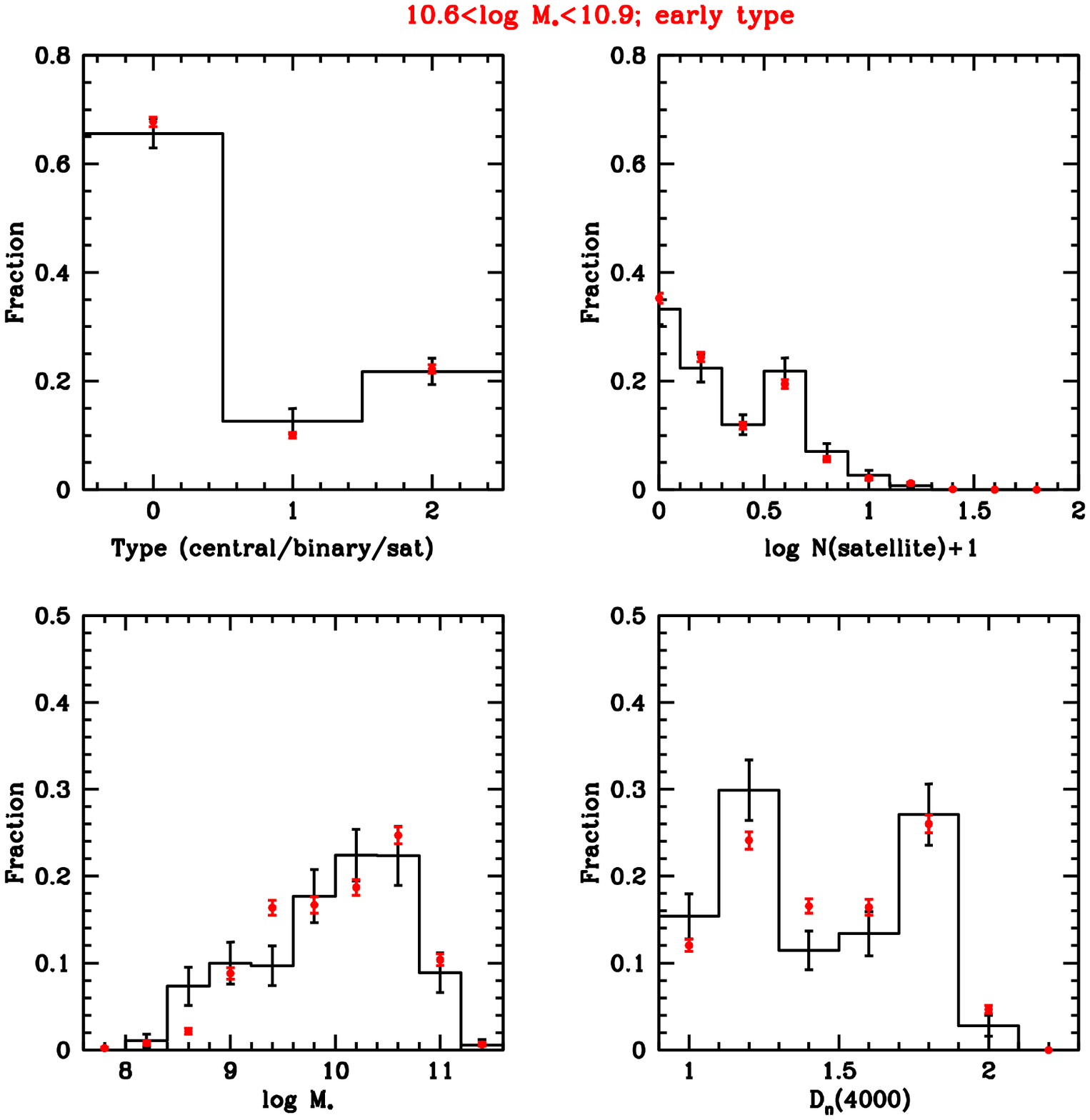}
\caption
{As in Figure 10, except for                                            
early-type ($C>2.6$) galaxies in the stellar mass range $10.6<\log M_*<10.9$.
\label{models}}
\end{figure}
 
\begin{figure}
\includegraphics[width=88mm]{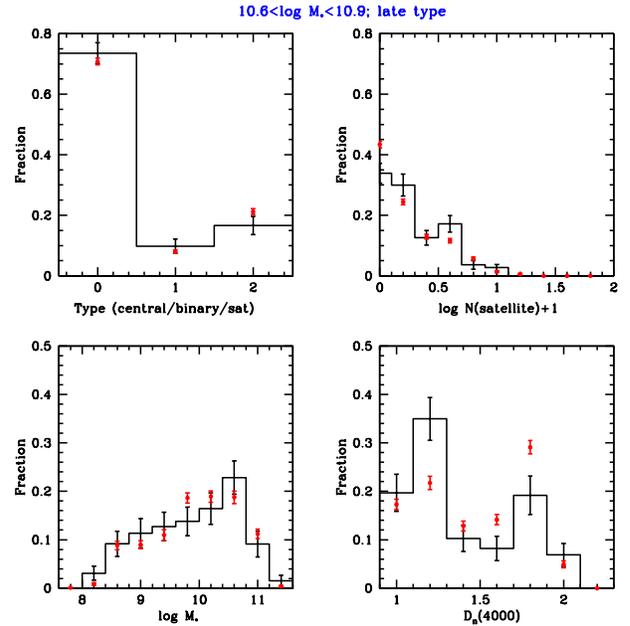}
\caption
{As in Figure 10, except for                                            
late-type ($C<2.6$) galaxies in the stellar mass range $10.6<\log M_*<10.9$.
\label{models}}
\end{figure}
 
\subsection {Holmberg effect}

In 1969, Holmberg wrote a paper claiming that satellite galaxies were
preferentially located along the minor axis of disk galaxies. Zaritsky et al
(1997) found a similar result for a sample of isolated spirals , but more recent
studies have led to different conclusions.  The most recent studies using large samples
drawn from redshift surveys such as 2dF and SDSS find that 
satellite alignments occur along the major
axis of the primary (Brainerd 2005; Yang et al 2006). In addition, alignments
are are  strongest for massive red galaxies (Yang et al 2006;
Faltenbacher et al 2007, 2009).  The standard explanation for satellite alignment
effects is that they occur as a result of the large-scale tidal field and the
preferred accretion of matter along filaments in the large-scale distribution of
galaxies.  Theoretical support for this scenario comes from studying the origin
of alignment effects between dark matter halos in N-body simulations
(Li et al 2013).  The most massive red galaxies are predicted to
be located  at the centers of the most massive dark matter halos, where
accretion of stellar material from infalling satellites occurs frequently, even
at the present day (e.g. Oser et al 2010). Alignment effects are thus expected to
be strong for these systems.

It is thus interesting to consider whether alignments can be used as a probe of
gas accretion processes.
Accretion of new gas
onto galaxies is thought to occur in two main ``modes": a) a so-called cold
mode, where new gas is accreted along filaments that trace the large-scale
galaxy distribution, b) a so-called hot mode, where new gas is accreted from a
spherically symmetric halo of gas in hydro-static equilibrium with the
surrounding halo (White \& Frenk 1991; Keres et al 2005). 
The theoretical expectation is that low mass galaxies in low
mass dark matter halos accrete gas in the cold mode, while high mass galaxies in
high mass halos accrete gas in the hot mode.

In this section, we study whether the satellite galaxies of HI-rich galaxies
exhibit preferred alignments with respect to the primary. 
We construct the distribution of the angle $\Theta$ between
the major axis of the primary and the line connecting the primary to the
satellite. $\Theta$ thus ranges from 0 to 90 degrees, with $\Theta=0$
corresponding to alignment along the major axis.  In Figure 13, we plot the
cumulative distribution of $\Theta$ for early and late-type HI-rich primaries in
two different mass ranges and compare this to the distribution obtained for the
control samples. We find that only the satellites of HI-rich galaxies in the low
stellar mass  bin($10.0 < \log M_* < 10.6$) exhibit a significant major axis
alignment effects. All the other HI-rich samples and all four control samples
yield a null result \footnote {The fact that we get a null result for the massive
early-type galaxy samples is not surprising, given that we have imposed a stringent
isoluation criterion on these systems, thereby elimating the majority of
galaxies in the most massive dark matter halos.} . We note that we have experimented with changing the
radius and the velocity difference within which a galaxy must fall to be
included as a satellite, and we find that alignment effects get stronger for
smaller values of $\Delta R$ and $\Delta V$. The results in Figure 13 show
results for $\Delta R <$ 250 kpc   and $\Delta V <$ 500 km/s, which represents a
compromise between the strength of the signal and the number of satellites
included in the sample.

\begin{figure}
\includegraphics[width=88mm]{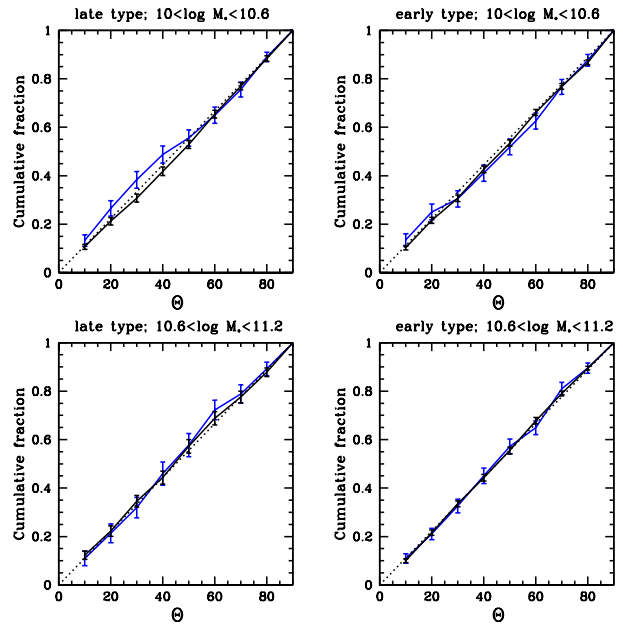}
\caption
{The cumulative fraction of $\Theta$, the angle between the major axis of a galaxy
and the line connecting the galaxy to one of its satellites. Results are plotted
for early and late-type galaxies in two different bins in stellar mass. Blue
lines show results for HI-rich galaxies in the upper 10th percentile in HI gas
mass fraction, while black lines show results for the control sample.
Error bars are computed by boot-strapping.
\label{models}}
\end{figure}

\section {Summary and Conclusions}
In this paper, we have investigated the nature of HI-rich galaxies in more detail than
in previous published work by analyzing their outer stellar populations using 
optical $g-r$ versus $r-z$  colour/colour diagrams, and by studying spectroscopically
identified neighbours within a radius of 500 kpc. All our results are compared with those
from control galaxy samples that are defined without regard to 
atomic gas content, but are matched in stellar mass, redshift, and structural 
parameters such as concentration and size. 
We restrict the analysis of outer colours to galaxies with axial ratios $b/a > 0.6$ and 
and we study HI-rich early-type ($C>2.6$) and late-type ($C<2.6$) galaxies separately. 

The outer stellar populations of HI-rich early-type galaxies  are shifted
with respect to control samples along
a locus in the colour/colour plane consistent with younger stellar ages, but similar  metallicities.
The outer stellar populations of HI-rich late-types are shifted much more to the blue 
in the $r-z$ direction, and we infer that they have outer disks which are both younger
and more metal-poor. The most extreme of these systems have outer disks with inferred metallicities
below one tenth solar. The outer colour shifts are stronger for HI-rich galaxies with lower masses.  
No outer colour shifts are seen for  HI-rich early-type  galaxies with stellar masses greater
than $ 3 \times 10^{10} M_{\odot}$. This may indicate that the HI disks of these systems form stars
less efficiently (so-called ``morphological quenching''; Martig et al 2009).

We also analyze the galaxy environments of HI-rich galaxies on scales comparable to
the expected virial radii of their dark matter halos ($R < 500$ kpc , $\Delta V < 500$ km/s).
Low mass ($\log M_* < 10.5$)
HI-rich early-type galaxies have galaxy environments that differ very significantly
from the control sample. HI-rich early types are more likely to be central
rather than satellite systems. Their satellites are also less massive and have
younger stellar populations. Similar, but weaker effects are found for
HI-rich late-type galaxies of the same mass. In addition, we find that the satellites of HI-rich late-types
exhibit a greater tendency to be aligned along the major axis of the primary.
No environmental differences are found for massive ($\log M_* > 10.5$) HI-rich galaxies,
regardless of type.   

Our main motivation for studying HI-rich galaxies is that they might serve as laboratories
for studying ongoing galaxy growth at the present day. The most important new insight from the
current study is that the way in which the build-up of stellar mass in the outer
regions of the galaxy is manifested, depends on both galaxy morphology and mass. In addition, the association of
HI-richness with differing galaxy environments for  low mass systems, but not
for high mass systems, hints that different accretion mechanisms may be at work
in galaxies of different mass, as predicted by currently popular
models of galaxy formation.  

We caution once again that broad-band photometry is a rather blunt tool in the
analysis of stellar populations. Much more can be learned if spectroscopy is available.
The effect of bursts on stellar age determinations  can be accounted for  using the combination of the 4000 \AA\ break strength
and Balmer absorption lines (Kauffmann et al 2003, 2014). Because these stellar age indicators span a very narrow 
wavelength range, they are relatively insensitive to the effects of dust. Finally, it is also possible to
constrain not only metallicity, but also element abundance ratios 
(Thomas, Maraston \& Bender  2003; Conroy, Graves \& Van Dokkum 2014) and
possibly the stellar  initial mass function. Next generation 
integral field unit (IFU) spectroscopic surveys  of tens of thousands of galaxies will
enable statistical studies of the spectroscopic properties of the  outer stellar populations
of galaxies for the first time. Next generation wide-field HI surveys  will 
obtained resolved maps of the atomic gas in galaxies, allowing us to understand the
processes governing star formation and chemical enrichment in disks in considerably more detail.



\end{document}